\documentclass[journal]{IEEEtran}

\usepackage{cite}
\usepackage[cmex10]{amsmath}
\usepackage{algorithmic}
\usepackage{amsfonts}
\usepackage{amssymb}
\usepackage{mathrsfs}
\usepackage{pgf}
\usepackage{tikz}
\usetikzlibrary{arrows,automata}
\usepackage[latin1]{inputenc}
\usepackage{verbatim}

\newtheorem{thm}{Theorem}

\newtheorem{lemma}{Lemma}

\newtheorem{definition}{\emph{Definition}}

\newcommand{\Rank}{{\mathrm{Rank}}}

\begin{document}
%
\title{On Random Linear Network Coding for Butterfly Network
\thanks{This research is supported in part by the National Natural Science Foundation of
China under the Grants 60872025 and 10990011. }}

\author{Xuan~Guang\thanks{X. Guang is with the Chern Institute of
Mathematics, Nankai University, Tianjin 300071, P.R. China. Email:
xuanguang@mail.nankai.edu.cn}, and~Fang-Wei~Fu\thanks{F.-W. Fu is with the Chern Institute of
Mathematics and LPMC, Nankai University, Tianjin 300071, P.R. China. Email:
fwfu@nankai.edu.cn}}

\maketitle
\begin{abstract}
Random linear network coding is a feasible encoding tool for network coding, specially for the non-coherent network, and its performance is important in theory and application. In this letter, we study the performance of random linear network coding for the well-known butterfly network by analyzing the failure probabilities. We determine the failure probabilities of random linear network coding for the well-known butterfly network and the butterfly network with channel failure probability $p$.
\end{abstract}

%
\IEEEpeerreviewmaketitle
\section{Introduction}
\IEEEPARstart{I}{n} the seminal paper \cite{Ahlswede-Cai-Li-Yeung-2000},  Ahlswede \emph{et
al} showed that with network
coding, the source node can multicast information to all sink
nodes at the theoretically maximum rate as the alphabet size
approaches infinity, where the theoretically maximum rate is the smallest minimum cut
capacity between the source node and any sink node. Li \emph{et al} \cite{Li-Yeung-Cai-2003}
showed that linear network coding with finite alphabet is
sufficient for multicast. Moreover, the well-known butterfly network was proposed in papers \cite{Ahlswede-Cai-Li-Yeung-2000}\cite{Li-Yeung-Cai-2003} to show the advantages of network coding well compared with routing. Koetter and M$\acute{\textup{e}}$dard
\cite{Koetter-Medard-algebraic} presented an algebraic
characterization of network coding. Ho \emph{et
al} \cite{Ho-etc-random} proposed the random linear network coding and gave several upper bounds on the failure probabilities of random linear network coding. Balli, Yan, and Zhang \cite{zhang-random} improved on these bounds and discussed the limit behavior of
the failure probability as the field size goes to infinity.

\begin{figure}[!h]
\begin{center}
\begin{tikzpicture}[->,>=stealth',shorten >=1pt,auto,node distance=1.2cm,
                    thick]
  \tikzstyle{every state}=[fill=none,draw=black,text=black,minimum size=7mm]
  \tikzstyle{place}=[fill=none,draw=white,minimum size=0.1mm]

  \node[state]         (s)                    {$s$};
  \node[place]         (o1) [above of=s,xshift=-3mm]            {};
  \node[place]         (o2) [above of=s,xshift=3mm]             {};
  \node[state]         (s1) [below left of=s,xshift=-10mm] {$s_1$};
  \node[state]         (s2) [below right of=s,xshift=10mm] {$s_2$};
  \node[state]         (i) [below right of=s1,xshift=10mm] {$i$};
  \node[state]         (j) [below of=i,yshift=-2mm]       {$j$};
  \node[state]         (t1)[below left of=j,xshift=-10mm]   {$t_1$};
  \node[state]         (t2)[below right of=j,xshift=10mm]  {$t_2$};

  \path (o1)edge  [dashed]    node [swap]{$d_1$}(s)
        (o2)edge  [dashed]    node {$d_2$}(s)
        (s) edge              node [swap]{$e_1$} (s1)
            edge              node {$e_2$} (s2)
        (s1)edge              node {$e_3$} (t1)
            edge              node {$e_4$} (i)
        (s2)edge              node[swap] {$e_5$} (i)
            edge              node {$e_6$} (t2)
        (i) edge              node {$e_7$} (j)
        (j) edge              node {$e_8$} (t1)
        (j) edge              node[swap] {$e_9$} (t2);
\end{tikzpicture}
\caption{Butterfly Network where $d_1,d_2$ represent two imaginary incoming channels of source node $s$.}
\label{fig_bn}
\end{center}
\end{figure}
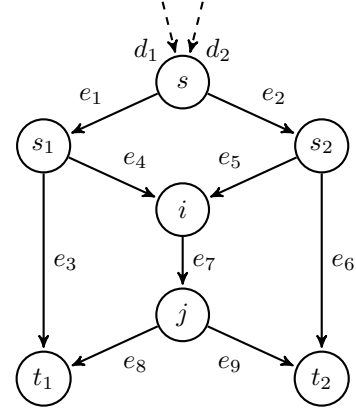

In this paper, we study the failure probabilities of random linear network coding for the well-known butterfly network when it is considered as a single source multicast network described by Fig. \ref{fig_bn}, and discuss the limit behaviors of
the failure probabilities as the field size goes to infinity.

\section{Failure Probabilities of Random Linear Network Coding for Butterfly Network}
In this letter, we always denote $G=(V,E)$ the butterfly network shown by Fig. \ref{fig_bn}, where the single source node is $s$, the set of sink nodes is $T=\{t_1,t_2\}$, the set of internal nodes is $J=\{s_1,s_2,i,j\}$, and the set of the channels is $E=\{e_1,e_2,e_3,e_4,e_5,e_6,e_7,e_8,e_9\}$.
Moreover, we assume that each channel $e\in E$ is error-free, and the capacity is unit 1.
It is obvious that both values of the maximum flows for the sink nodes $t_1$ and $t_2$ are 2.

  Now we consider random linear network coding problem with the information rate 2 for this butterfly network $G$. That is to say, for source node $s$ and internal nodes, all local encoding coefficients are independently uniformly distributed random variables taking values in the base field $\mathcal{F}$. For source node $s$, although it has no incoming channels, we use two imaginary incoming channels $d_1, d_2$ and assume that they provide the source messages for $s$. Let the global encoding kernels of
$f_{d_1}=\begin{pmatrix}
1\\
0
\end{pmatrix}$
and
$f_{d_2}=\begin{pmatrix}
0\\
1
\end{pmatrix}$,
respectively. The local encoding kernel at the source node $s$ is denoted as $K_s=\begin{pmatrix}k_{d_1,1}&d_{d_1,2}\\k_{d_2,1}&d_{d_2,2}\end{pmatrix}$.

Similarly, we denote by $f_i$ the global encoding kernel of channel $e_i\ (1\leq i \leq 9)$, and $k_{i,j}$ the local encoding coefficient for the pair $(e_i,e_j)$ with $head(e_i)=tail(e_j)$ as described in \cite[p.442]{Yeung-book}, where $tail(e_i)$ represents the node whose outgoing channels include $e_i$, $head(e_i)$ represents the node whose incoming channels include $e_i$:
$$f_1=k_{d_1,1}f_{d_1}+k_{d_2,1}f_{d_2}=
\begin{pmatrix}
k_{d_1,1}\\
k_{d_2,1}
\end{pmatrix},$$
$$f_2=k_{d_1,2}f_{d_1}+k_{d_2,2}f_{d_2}=
\begin{pmatrix}
k_{d_1,2}\\
k_{d_2,2}
\end{pmatrix},$$
$$f_3=k_{1,3}f_1=\begin{pmatrix}
k_{d_1,1}k_{1,3}\\
k_{d_2,1}k_{1,3}
\end{pmatrix}
,
f_4=k_{1,4}f_1=\begin{pmatrix}
k_{d_1,1}k_{1,4}\\
k_{d_2,1}k_{1,4}
\end{pmatrix},$$
$$f_5=k_{2,5}f_2=
\begin{pmatrix}
k_{d_1,2}k_{2,5}\\
k_{d_2,2}k_{2,5}
\end{pmatrix}
,
f_6=k_{2,6}f_2=\begin{pmatrix}
k_{d_1,2}k_{2,6}\\
k_{d_2,2}k_{2,6}
\end{pmatrix}
,$$
$$f_7=k_{4,7}f_4+k_{5,7}f_5=
\begin{pmatrix}
k_{d_1,1}k_{1,4}k_{4,7}+k_{d_1,2}k_{2,5}k_{5,7}\\
k_{d_2,1}k_{1,4}k_{4,7}+k_{d_2,2}k_{2,5}k_{5,7}
\end{pmatrix}
,$$
$$f_8=k_{7,8}f_7=
\begin{pmatrix}
k_{d_1,1}k_{1,4}k_{4,7}k_{7,8}+k_{d_1,2}k_{2,5}k_{5,7}k_{7,8}\\
k_{d_2,1}k_{1,4}k_{4,7}k_{7,8}+k_{d_2,2}k_{2,5}k_{5,7}k_{7,8}
\end{pmatrix}
,$$
$$f_9=k_{7,9}f_7=
\begin{pmatrix}
k_{d_1,1}k_{1,4}k_{4,7}k_{7,9}+k_{d_1,2}k_{2,5}k_{5,7}k_{7,9}\\
k_{d_2,1}k_{1,4}k_{4,7}k_{7,9}+k_{d_2,2}k_{2,5}k_{5,7}k_{7,9}
\end{pmatrix}$$
and the decoding matrices of the sink nodes $t_1,t_2$ are
\begin{gather}
F_{t_1}=\begin{pmatrix}
k_{d_1,1}k_{1,3}&k_{d_1,1}k_{1,4}k_{4,7}k_{7,8}+k_{d_1,2}k_{2,5}k_{5,7}k_{7,8}\\
k_{d_2,1}k_{1,3}&k_{d_2,1}k_{1,4}k_{4,7}k_{7,8}+k_{d_2,2}k_{2,5}k_{5,7}k_{7,8}
\end{pmatrix},\label{Ft1}\\
F_{t_2}=\begin{pmatrix}
k_{d_1,2}k_{2,6}&k_{d_1,1}k_{1,4}k_{4,7}k_{7,9}+k_{d_1,2}k_{2,5}k_{5,7}k_{7,9}\\
k_{d_2,2}k_{2,6}&k_{d_2,1}k_{1,4}k_{4,7}k_{7,9}+k_{d_2,2}k_{2,5}k_{5,7}k_{7,9}\end{pmatrix}.\label{Ft2}
\end{gather}
\subsection{Butterfly Network without Fail Channels}
Next, we begin to discuss the failure probabilities of random linear network coding for the butterfly network $G$. These failure probabilities can illustrate the performance of the random linear network coding for the butterfly network. First, we give the definitions of failure probabilities.
\begin{definition}
For random linear network coding for the butterfly network $G$ with information rate $2$:
\begin{itemize}
    \item $P_{e}(t_i)\triangleq Pr(\Rank(F_{t_i})<2)$ is called the failure probability of sink node $t_i$, that is the probability that the messages cannot be decoded correctly at $t_i\ (i=1,2)$;
    \item $P_e\triangleq Pr(\exists\ t\in T \;\mbox{such that}\; \Rank(F_t)<2)$ is called
the failure probability of the butterfly network $G$, that is the probability that
the messages cannot be decoded correctly at at least one sink node
in $T$;
    \item $P_{av}\triangleq \frac{\sum_{t\in T}P_{e}(t)}{|T|}$ is called the average failure probability of all sink nodes for the butterfly network $G$.
\end{itemize}
\end{definition}
To determine the failure probabilities of random linear network coding for the butterfly network $G$, we need the following lemma.
\begin{lemma}\label{lem_matrix}
Let $M$ be a uniformly distributed random $2\times 2$ matrix over a finite field $\mathcal{F}$, then the probability that $M$ is invertible is $(|\mathcal{F}|+1)(|\mathcal{F}|-1)^2/{|\mathcal{F}|^3}.$
\end{lemma}
\begin{thm}\label{prob_without_fail}
For random linear network coding for the butterfly network $G$ with information rate $2$:
\begin{itemize}
  \item the failure probability of the sink node $t_i\ (i=1,2)$ is
        $$P_{e}(t_i)=1-\frac{(|\mathcal{F}|+1)(|\mathcal{F}|-1)^6}{|\mathcal{F}|^7}\ ;$$
  \item the failure probability of the butterfly network $G$ is
        $$P_e=1-\frac{(|\mathcal{F}|+1)(|\mathcal{F}|-1)^{10}}{|\mathcal{F}|^{11}}\ ;$$
  \item the average failure probability is
        $$P_{av}=1-\frac{(|\mathcal{F}|+1)(|\mathcal{F}|-1)^6}{|\mathcal{F}|^7}\ .$$
\end{itemize}
\end{thm}
\begin{proof}
Recall that $K_s=(f_1\ f_2)$ is the local encoding kernel at the source node $s$. Denote
\begin{equation*}
\begin{split}
B_1&\triangleq\begin{pmatrix}
k_{1,3}&k_{1,4}k_{4,7}k_{7,8}\\0&k_{2,5}k_{5,7}k_{7,8}\end{pmatrix},B_2\triangleq\begin{pmatrix}
0&k_{1,4}k_{4,7}k_{7,9}\\k_{2,6}&k_{2,5}k_{5,7}k_{7,9}\end{pmatrix}.
\end{split}
\end{equation*}
Then, we know from (\ref{Ft1}) and (\ref{Ft2}) that $F_{t_1}=K_sB_1,\ F_{t_2}=K_sB_2$.
Moreover, from Lemma \ref{lem_matrix}, we have
\begin{equation*}
\begin{split}
1-P_{e}(t_1)&=Pr(\Rank(F_{t_1})=2)=Pr(\det(F_{t_1})\neq 0)\\
&=Pr(\{\det(K_s)\neq0\}\cap\{\det(B_1)\neq 0\})\\
&=Pr(\det(K_s)\neq0)\cdot Pr(\det(B_1)\neq 0)\\
&=\frac{(|\mathcal{F}|+1)(|\mathcal{F}|-1)^2}{|\mathcal{F}|^3}
\left(\frac{|\mathcal{F}|-1}{|\mathcal{F}|}\right)^4\\
&=\frac{(|\mathcal{F}|+1)(|\mathcal{F}|-1)^6}{|\mathcal{F}|^7}.
\end{split}
\end{equation*}
Similarly, we also have $1-P_{e}(t_2)=(|\mathcal{F}|+1)(|\mathcal{F}|-1)^6/{|\mathcal{F}|^7}$.

Next, we consider the failure probability $P_e$. Similar to the above analysis, we have
\begin{equation*}
\begin{split}
&1-P_e=Pr(\{\Rank(F_{t_1})=2\} \cap \{\Rank(F_{t_2})=2\})\\
=&Pr(\{\det(F_{t_1})\neq 0\} \cap \{\det(F_{t_2})\neq 0\})\\
=&Pr(\{\det(K_s)\neq0\}\cap\{\det(B_1)\neq 0\}\cap\{\det(B_2)\neq 0\})\\
=&Pr(\det(K_s)\neq0)\cdot Pr(\det(B_1)\neq 0)\cdot Pr(\det(B_2)\neq 0)\\
=&\frac{(|\mathcal{F}|+1)(|\mathcal{F}|-1)^{10}}{|\mathcal{F}|^{11}}.\\
\end{split}
\end{equation*}
Moreover, since $P_{av}=\frac{P_{e}(t_1)+P_{e}(t_2)}{2}$, then we have
\begin{equation*}
P_{av}=1-\frac{(|\mathcal{F}|+1)(|\mathcal{F}|-1)^6}{|\mathcal{F}|^7}\ .
\end{equation*}
This completes the proof.
\end{proof}

It is known that if $|\mathcal{F}| \geq |T|$, there exists a linear network code
$\mathbf{C}$ such that all sink nodes can decode successfully $w=\min_{t\in T}C_t$ symbols generated by the source node $s$, where $C_t$ is the minimum cut capacity between $s$ and the sink node $t$. For butterfly network, $|\mathcal{F}|\geq 2$ is enough. However, when we consider the performance of random linear network coding, the result is different.

For $|\mathcal{F}|=2$, we have $1-P_{e}=\frac{3}{2^{11}}\approx 0.001$. In other words, this successful probability is too much low for application. In the same way, for $|\mathcal{F}|=3$ and $|\mathcal{F}|=4$, the successful probabilities are approximately $0.023$ and $0.070$, respectively. These successful probabilities are still too low for application.

If $1-P_e \geq 0.9$ is thought of enough for application, we have to choose a very large base field $\mathcal{F}$ with cardinality not less than 87. This is because
$$\frac{(|\mathcal{F}|+1)(|\mathcal{F}|-1)^{10}}{|\mathcal{F}|^{11}}\geq 0.9 \Longleftrightarrow |\mathcal{F}|\geq 87\ . $$


\subsection{Butterfly Network with Fail Channels}
In this subsection, as in \cite{Ho-etc-random}, we will consider the failure probabilities of random linear network coding for the butterfly network with channel failure probability $p$ (i.e., each channel is possibly deleted from the network with probability $p$). Generally speaking, channel failure is a low probability event, i.e., $0\leq p \ll \frac{1}{2}$. In fact, if $p=0$, it is just the butterfly network without fail channels discussed in the above subsection.

\begin{thm}\label{prob_with_fail}
For the random linear network coding of the butterfly network $G$ with channel failure probability $p$,
\begin{itemize}
  \item the failure probability of the sink node $t_i\ (i=1,2)$ is
        $$\tilde{P}_{e}(t_i)=1-\frac{(|\mathcal{F}|+1)(|\mathcal{F}|-1)^6}{|\mathcal{F}|^7}(1-p)^6\ ;$$
  \item the failure probability of the butterfly network $G$ is
        $$\tilde{P}_e=1-\frac{(|\mathcal{F}|+1)(|\mathcal{F}|-1)^{10}}{|\mathcal{F}|^{11}}(1-p)^9\ ;$$
  \item the average failure probability is
        $$\tilde{P}_{av}=1-\frac{(|\mathcal{F}|+1)(|\mathcal{F}|-1)^6}{|\mathcal{F}|^7}(1-p)^6\ .$$
\end{itemize}
\end{thm}
\begin{proof}
At first, we consider the failure probability $\tilde{P}_{e}(t_1)$. For each channel $e_i$, we call ``channel $e_i$ is successful'' if $e_i$ is not deleted from the network, and define $\delta_i$ as the event that ``$e_i$ is successful''. We define $\tilde{f_i}$ as the \textit{active global encoding kernel} of $e_i$, where
$$\tilde{f}_i=\left\{\begin{array}{ll}f_i,\ &\mbox{$e_i$ is successful,}\\
                             \underline{0}\ ,\ &\mbox{otherwise.}\end{array}\right.$$
and $f_i=\sum_{j: head(e_j)=tail(e_i)}k_{j,i}\tilde{f}_j$. Then the decoding matrix of the sink node $t_1$ is $F_{t_1}=(\tilde{f}_3\ \tilde{f}_8)$. Hence, $\tilde{P}_{e}(t_1)=Pr(\Rank((\tilde{f}_3\ \tilde{f}_8))\neq2)$.

Note that the event ``$\Rank((\tilde{f}_3\ \tilde{f}_8))=2$'' is equivalent to the event ``$\Rank((f_3\ f_8))=2$, $\delta_3$, and $\delta_8$''. Moreover, since $(f_3\ f_8)=(k_{1,3}\tilde{f}_1\ k_{7,8}\tilde{f}_7)$, we have $\Rank((f_3\ f_8))=2$ if and only if $\Rank((\tilde{f}_1\ \tilde{f}_7))=2$ and $k_{1,3}\neq0, k_{7,8}\neq0$. Similarly, the event ``$\Rank((\tilde{f}_1\ \tilde{f}_7))=2$'' is equivalent to the event ``$\Rank((f_1\ f_7))=2$, $\delta_1$, and $\delta_7$''. Note that $\tilde{f}_4=\underline{0}$ or $f_4$, and $f_4=k_{1,4}\tilde{f}_1$, so $\tilde{f}_4=\underline{0}$ or
$k_{1,4}f_1$. Since $f_7=k_{4,7}\tilde{f}_4+k_{5,7}\tilde{f}_5$, we have $\Rank((f_1\ f_7))=2$ if and only if $\Rank((f_1\ \tilde{f}_5))=2$ and $k_{5,7}\neq0$. Going on with the same analysis as above, we have that the event ``$\Rank((f_1\ \tilde{f_5}))=2$'' is equivalent to the event
``$\Rank((f_1\ f_5))=2$, and $\delta_5$'', and the event ``$\Rank((f_1\ f_5))=2$'' is equivalent to the event ``$\Rank((f_1\ f_2))=2$, $k_{2,5}\neq0$, and $\delta_2$'' since $f_5=k_{2,5}\tilde{f}_2$. Therefore, the event ``$\Rank((\tilde{f}_3\ \tilde{f}_8))=2$'' is equivalent to the event ``$\Rank((f_1\ f_2))=2$, $\delta_1$, $\delta_2$, $\delta_3$, $\delta_5$, $\delta_7$, $\delta_8$, and $k_{1,3},k_{2,5},k_{5,7},k_{7,8}\neq0$".
Hence, we have
\begin{equation*}
\begin{split}
&1-\tilde{P}_{e}(t_1)=Pr(\Rank((\tilde{f}_3\ \tilde{f}_8))=2)\\
=&Pr(\{\Rank(f_1\ f_2)=2\}\cap \{\delta_1,\delta_2,\delta_3,\delta_5,\delta_7,\delta_8\}\\
&\ \ \ \ \ \cap\{k_{1,3},k_{2,5},k_{5,7},k_{7,8}\neq0\})\\
=&Pr(\{\Rank(f_1\ f_2)=2\})\cdot Pr(\{\delta_1,\delta_2,\delta_3,\delta_5,\delta_7,\delta_8\})\\
&\ \ \ \ \ \cdot Pr(\{k_{1,3},k_{2,5},k_{5,7},k_{7,8}\neq0\})\\
=&\frac{(|\mathcal{F}|+1)(|\mathcal{F}|-1)^6}{|\mathcal{F}|^7}(1-p)^6.\\
\end{split}
\end{equation*}

In the same way, we can get $$\tilde{P}_{e}(t_2)=\tilde{P}_{e}(t_1)=1-\frac{(|\mathcal{F}|+1)(|\mathcal{F}|-1)^6}{|\mathcal{F}|^7}(1-p)^6\ .$$

For the failure probability $\tilde{P}_e$, with the same analysis as above, we have
\begin{equation*}
\begin{split}
&1-\tilde{P}_e=Pr(\{\Rank(F_{t_1})=2\}\cap \{\Rank(F_{t_2})=2\})\\
=&Pr(\{\Rank((\tilde{f}_3\ \tilde{f}_8))=2\}\cap\{\Rank((\tilde{f}_6\ \tilde{f}_9))=2\})\\
\end{split}
\end{equation*}
\begin{equation*}
\begin{split}
=&Pr(\{\Rank((f_1\ f_2))=2\}\cap\{\delta_1,\delta_2,\delta_3,\delta_4,\delta_5,\delta_6,\delta_7,\delta_8,\delta_9\}\\
&\cap\{k_{1,3},k_{2,5},k_{5,7},k_{7,8},k_{2,6},k_{1,4},k_{4,7},k_{7,9}\neq 0\})\\
=&Pr(\Rank((f_1\ f_2))=2)\cdot Pr(\delta_1,\delta_2,\delta_3,\delta_4,\delta_5,\delta_6,\delta_7,\delta_8,\delta_9)\\
&\cdot Pr(\{k_{1,3},k_{2,5},k_{5,7},k_{7,8},k_{2,6},k_{1,4},k_{4,7},k_{7,9}\neq 0\})\\
=&\frac{(|\mathcal{F}|+1)(|\mathcal{F}|-1)^{10}}{|\mathcal{F}|^{11}}(1-p)^9.
\end{split}
\end{equation*}
At last, the average probability $\tilde{P}_{av}$ is given by
$$\tilde{P}_{av}=\frac{\tilde{P}_{e}(t_1)+\tilde{P}_{e}(t_2)}{2}=1-\frac{(|\mathcal{F}|+1)(|\mathcal{F}|-1)^6}{|\mathcal{F}|^7}(1-p)^6\ .$$
The proof is completed.
\end{proof}


\section{The Limit Behaviors of The Failure Probabilities}
In this section, we will discuss the limit behaviors of the failure probabilities. From Theorem \ref{prob_with_fail}, we have
$$\begin{array}{ll}
\lim\limits_{|\mathcal{F}|\rightarrow \infty}\tilde{P}_{e}(t_i)=\lim\limits_{|\mathcal{F}|\rightarrow \infty}\tilde{P}_{av}=1-(1-p)^6\ ,(i=1,2)\ ,\\
\lim\limits_{|\mathcal{F}|\rightarrow \infty}\tilde{P}_e=1-(1-p)^9\ .
\end{array}$$
Specially, if we consider the random linear network coding for the butterfly network without fail channels, i.e. $p=0$, we have
(also from Theorem \ref{prob_without_fail})
$$
\lim_{|\mathcal{F}|\rightarrow \infty}P_{e}(t_i)=\lim_{|\mathcal{F}|\rightarrow \infty}P_e=
\lim_{|\mathcal{F}|\rightarrow \infty}P_{av}=0\ ,\ (i=1,2).$$
That is, for the sufficient large base field $\mathcal{F}$, the failure probabilities can become arbitrary small. In fact, this result is correct for all single source multicast network coding \cite{Ho-etc-random}.
It is easy to see from Theorem \ref{prob_without_fail} that the rates of $P_{e}(t_i)$ approaching 0 and $P_{av}$ approaching 0 are $\frac{5}{|\mathcal{F}|}$, and the rate of $P_{e}$ approaching 0 is $\frac{9}{|\mathcal{F}|}$.


\section{Conclusion}

The performance of random linear network coding is illustrated by the failure probabilities of random linear network coding. In general, it is very difficult to calculate the failure probabilities of random linear network coding for a general communication network. In this letter, we determine the failure probabilities of random linear network coding for the well-known butterfly network and the butterfly network with channel failure probability $p$.


\end{document}